\def\spose#1{\hbox to 0pt{#1\hss}}

\def\multleft#1{\hbox to size{\vbox {\halign {\lft{##}\cr #1}}\hfill}\par}
\def\multright#1{\hbox to size{\vbox {\halign {\rt{##}\cr #1}}\hfill}\par}

\def\today{\ifcase\month\or January\or February\or March\or April\or May\or
      June\or July\or August\or September\or October\or November\or December\fi
      \space\number\day, \number\year}

\def\H2{\hbox{H$_{2}$}}

\documentclass{mn2e}
\usepackage{times}
\usepackage{amssymb}
\usepackage{epsfig}
\usepackage{lscape}
\usepackage{graphicx}
\begin{document}
\hsize=6truein
          
\title[The $\bmath{z=9}$ galaxy population]
{New redshift $\bmath{z \simeq 9}$ galaxies in the Hubble Frontier Fields: Implications
for early evolution of the UV luminosity density}
\author[D.J.~McLeod et al.]
{D.J.\,McLeod$^{1}$\thanks{Email: mcleod@roe.ac.uk}, R.J.\,McLure$^{1}$,
  J.S.\,Dunlop$^{1}$, B.E.\,Robertson$^{2}$,  R.S.\,Ellis$^{3}$
and T.A.\,Targett$^{4}$
\footnotesize\\
$^{1}$SUPA\thanks{Scottish Universities Physics Alliance}, Institute
for Astronomy, University of Edinburgh, Royal Observatory, Edinburgh
EH9 3HJ\\
$^{2}$Department of Astronomy and Steward Observatory, University of Arizona, Tucson, AZ 85721, USA\\
$^{3}$Department of Astrophysics, California Institute of Technology, MS 249-17, Pasadena, CA 91125, USA\\
$^{4}$Department of Physics and Astronomy, Sonoma State University, Rohnert Park, CA 94928, USA}
\maketitle

\begin{abstract}
We present the results of a new search for galaxies at redshift
$z \simeq 9$ in the first two Hubble Frontier Fields with completed {\it HST} WFC3/IR and ACS imaging. To ensure robust photometric redshift solutions, 
and to minimize incompleteness, we confine our search to objects with $H_{160} < 28.6$\,(AB mag), consider only image regions with
an rms noise $\sigma_{160} > 30$\,mag (within a 0.5-arcsec diameter aperture), and 
insist on detections in both $H_{160}$ and $J_{140}$. The result is a survey covering an effective 
area (after accounting for magnification) of 10.9\,arcmin$^2$, 
which yields 12 galaxies at $8.4 < z < 9.5$. Within the Abell-2744
cluster and parallel fields we confirm the three brightest objects reported by Ishigaki et al.,
but recover only one of the four $z > 8.4$ sources reported by Zheng et al.. 
In the MACSJ0416.1-240 cluster field we report five objects, and explain why each of these eluded detection 
or classification as $z \simeq 9$ galaxies in the published searches of the shallower CLASH data.
Finally we uncover four $z \simeq 9$ galaxies from the MACSJ0416.1-240 parallel field. Based on the published magnification maps we find
that only one of these 12 galaxies is likely boosted by more than a factor of two by gravitational lensing. 
Consequently we are able to perform a fairly straightforward reanalysis of the normalization of the $z \simeq 9$ UV galaxy
luminosity function as explored previously in the HUDF12 programme. We conclude that the new 
data strengthen the evidence for a continued smooth decline in UV luminosity density (and hence 
star-formation rate density) from $z \simeq
8$ to $z \simeq 9$, contrary to recent reports of a marked drop-off at these redshifts. 
This provides further support for the scenario in which early galaxy evolution is sufficiently
extended to explain cosmic reionization.
\end{abstract}


\section{INTRODUCTION}

The discovery and study of galaxies at $z \ge 7$ was first made possible by the advent of deep, multi-band, near-infrared imaging provided by the installation of WFC3/IR on the 
{\it Hubble Space Telescope} ({\it HST}) in 2009 (e.g. Bouwens et al. 2010; McLure et al. 2010; 
Oesch et al. 2010; Finkelstein et al. 2010; Bunker et al. 2010; see Dunlop 2013 for a review). The initial deep studies with WFC3/IR revealed 
the first significant samples of galaxies at $z \simeq 7$ and $z \simeq 8$, based on imaging through 
the $Y_{105}$ (or $Y_{098}$), $J_{125}$ and $H_{160}$ filters, coupled with pre-existing deep 
optical imaging with ACS (to establish a lack of emission shortward of the redshifted 
Lyman-break at $\lambda_{rest} = 1216$\AA). In the intervening 5 years, by combining the 
ultra-deep data available in the Hubble Ultra Deep Field (HUDF; HUDF09+HUDF12) 
with shallower, wider-field data from the HUDF09 parallel fields, CANDELS (Grogin et
al. 2011; Koekemoer et al. 2011) and BoRG surveys (Trenti et
al. 2011), several different groups have reached very similar
conclusions on the form of the rest-frame UV galaxy luminosity 
function (LF) at $z \simeq 7$ and $z\simeq 8$ (e.g. McLure et
al. 2010, 2013; Oesch et al. 2012b; Bradley et
al. 2012; Schenker et al. 2013; Schmidt et al. 2014; Bouwens et al. 2015; Finkelstein et al. 2014),
at least at $M_{1500}>-21$ (see Bowler et al. 2012, 2014a, 2014b for new progress 
in the continuing 
debate over the form of the bright-end of the high-redshift galaxy LF).

While a general consensus has therefore emerged over the evolving LF (and hence UV luminosity 
density) out to $z \simeq 8$, progress at higher redshifts has, unsurprisingly, proved more difficult.
The robust selection of galaxies at $z\simeq 9$ required exploitation of
the additional $J_{140}$ filter, enabling deeper, dual-band 
($J_{140} + H_{160}$) detections of the rest-frame UV continuum
longward of the Lyman break even if the source lay beyond $z\simeq 8.5$.
This approach (coupled with ultra-deep $Y_{105}$ imaging) 
was utilised in the HUDF12 programme to provide the first meaningful sample of galaxies at $z
\simeq 9$ (Ellis et al. 2013; McLure et al. 2013) and was also successfully exploited
in the CLASH Treasury Programme to uncover examples of gravitationally-lensed galaxies at 
$z \simeq 9.6$ (Zheng et al. 2012) and $z \simeq 10.7$ (Coe et al. 2013).

Ellis et al. (2013) used the first robust sample of galaxies 
at $8.5<z<9.5$ uncovered from the HUDF12 imaging to investigate the
evolution of UV luminosity density at $z\geq 8$, and 
concluded that, although the star-formation rate density
at $z\geq 8$ appeared to fall below a simple extrapolation of the evolution seen
over $z \simeq 6 - 8$, the fall-off was less dramatic than implied by
some previous studies (i.e. Bouwens et al. 2011a; Oesch et al. 2012a),
and more consistent with estimates from individually lensed $z\geq 8$ galaxies from the
CLASH clusters survey (e.g. Zheng et al. 2012; Coe et al. 2013).
Subsequently, by combining the HUDF12 dataset with wider-area imaging from CANDELS, McLure et
al. (2013) produced the first constraints on the $z=9$ LF (at
$M_{1500}\simeq -18$) and used these to re-calculate the UV luminosity
density. The resulting inferred star-formation rate density at $z\simeq 9$ 
was actually somewhat higher than derived
by Ellis et al. (2013).

However, despite this progress, the relatively small size of the $z \simeq 9$ samples, 
coupled with the potential for extreme field-to-field cosmic variance when sampling 
only the bright end of the $z \simeq 9$ galaxy UV LF (e.g. Robertson et al. 2014), 
has led to considerable debate over the amount of star formation at $z \simeq 9$. 
In essence, while there is now a broad consensus that UV luminosity density declines smoothly and
gradually from $z \simeq 4$ to $z \simeq 8$, some authors have continued to argue that the available 
information at higher redshifts favours a rapid drop-off at $z \simeq 9$ 
(e.g. Oesch et al. 2013, 2014a, 2014b) while others have argued in support of continued smooth 
decline (e.g. Ellis et al. 2013).

This, then, is the new focus of debate in the study of high-redshift galaxy evolution. It is of importance for 
several reasons. First, since $z \simeq 9$ is only 95\,Myr earlier than $z \simeq 8$, a sudden
acceleration in the descent of UV luminosity density over such a relatively small amount of additional 
lookback time could be reasonably viewed as discovery of the `epoch of galaxy formation'. 

Secondly, such a down-turn in galaxy abundance would make it
more difficult for young galaxies to be the primary agent of cosmic
reionization (c.f. Robertson et al 2010, 2013), unless the optical depth 
due to Thomson scattering of the microwave background has been over-estimated
(Hinshaw et al 2013, Planck Collaboration 2014).

Third, such rapid evolution might also be viewed as unexpected given the known,
relatively mature properties of galaxies at $z \simeq 7$ (Dunlop et al. 2012, 2013) and theoretical predictions of early 
structure growth (e.g. Jaacks et al. 2012; Kimm \& Cen 2013; 
Paardekooper et al. 2013; Cai et al. 2014;  
Dayal et al. 2014; Genel et al. 2014; Henriques et al. 2014). Fourth, the evolution of the galaxy population out 
to $z \simeq 9$ has obvious implications for the prospects of the search for and study of galaxies at 
even earlier times (potentially out to $ z\simeq 20$) with the {\it James Webb Space Telescope} ({\it JWST}) from late 2018.

It is clear, therefore, that improved infomation on the prevalence of galaxies at $z \simeq 9$ is
required, and thankfully, due to the continued excellent performance of WFC3/IR, this is an 
area in which the {\it HST} can still make an important contribution. This is one of the science drivers for
the Hubble Frontier Fields (HFF) programme\footnote{http://www.stsci.edu/hst/campaigns/frontier-fields/}, 
which aims to deliver deep, near-infrared ($Y_{105}$, $J_{125}$, $J_{140}$, $H_{160}$) data 
(with associated ACS optical imaging), over 12 WFC3/IR pointings, consisting of 6 cluster and 
6 parallel fields. Imaging of the first cluster field, Abell-2744, was completed in early 2014 
and has been explored by several authors (Zheng et al. 2014; 
Zitrin et al. 2014; 
Ishigaki et al. 2015; Atek et al. 2015; Oesch et al. 2014b), 
while the Abell-2744 parallel field has also been interrogated for high-redshift galaxies (Ishigaki et al. 2015; 
Oesch et al. 2014b). At the time of writing, WFC3/IR+ACS imaging of the second cluster field, MACSJ0416.1-240
(previously targeted in the CLASH programme; Bradley et al. 2014; Bouwens et al. 2014),
and its associated parallel field has now also been completed.

While not of the depth achieved in the HUDF (see Koekemoer et al. 2013), these new pointings provide a large and important increase 
in the sky area covered with the crucial deep 4-band near-infrared imaging (i.e. including $J_{140}$)
required to robustly select galaxies at $z > 8.5$, and several of the aforementioned studies have reported 
galaxies in the Abell-2744 cluster and parallel fields at these extreme redshifts. Indeed, given the continued degradation
of the ACS, and relatively shallow nature of the HFF ACS optical imaging compared 
to the associated WFC3/IR imaging (at least compared to the HUDF) the HFF programme is arguably better 
optimized for the selection of galaxies at $z \simeq 9$ than at $z \simeq 7$.
Moreover, while much attention has been focussed on the potential for gravitional lensing in the cluster 
fields to yield boosted examples of faint high-redshift galaxies beyond the limits of blank field surveys 
such as the HUDF, it is the parallel fields that in fact provide the most valuable boost in 
well-understood cosmic volume for improving our knowledge of the number density of galaxies at $z \simeq 9$.

In this paper we exploit the new HFF data from all four of the pointings completed to date to photometrically 
select a new, robust sample of galaxies in the redshift range $8.4 < z < 9.5$. We apply the tried and
trusted methods developed through the HUDF12 programme (McLure et al. 2013), confining attention to galaxies which are 
sufficiently bright (in practice $H_{160} < 28.6$\,AB mag) for the available dynamic range in the
multi-colour photometry to deliver reasonably robust photometric redshifts. This approach 
also involves confining our attention to clean, and deep regions of the imaging
(where the local rms noise in $H_{160}$ as measured from $\simeq 150$ nearby 0.5-arcsec diameter apertures
is $\sigma_{160} \ge 30$\,mag), 
resulting in avoidance of the inner cluster regions, and most of the area 
traversed by the critical lines of high magnification. The outcome
is essentially a new, blank-field survey for galaxies at $z \simeq 9$, which complements 
the HUDF12 project by sampling $\simeq 1$\,mag shallower, but over an effective area $\simeq 3$ 
times greater. We use our new sample of $z \simeq 9$ galaxies to revisit our previous estimate of the UV LF (and hence luminosity density) at these redshifts, and briefly
explore the implications for early galaxy evolution.

The structure of the paper is as follows. In Section 2 we discuss the HFF imaging data and 
briefly describe our selection methods. In Section 3 we present and discuss our final sample of 
$z\simeq 9$ galaxies. 
In Section 4 we update our determination of the $z \simeq 9$ LF,  
derive the corresponding improved estimate of the integrated UV luminosity density 
and hence star-formation density at these redshifts, and compare with theoretical predictons. 
Finally, our conclusions are summarized  
in Section 5. Throughout the paper we will refer to the following {\it HST} ACS+WFC3/IR filters: 
F435W, F600LP, F606W, F775W, F814W, F850LP, F098M, F105W, F110W, F125W, F140W, F160W as 
$B_{435}, V_{600}, V_{606}, i_{775}, i_{814}, z_{850}, Y_{098}, Y_{105}, J_{110}, J_{125}, J_{140}, H_{160}$ respectively.
All magnitudes are quoted in the AB system (Oke 1974; Oke \& Gunn
1983) and all cosmological calculations 
assume $\Omega_{M}=0.3, \Omega_{\Lambda}=0.7$ and $H_{0}=70$ kms$^{-1}$Mpc$^{-1}$.

\section{Data and Object selection}

\subsection{HST, VLT, Gemini, and Spitzer Imaging}

The core dataset for this study is supplied by the 
public {\it HST} WFC3/IR+ACS imaging which has now been completed for the 
first two HFF clusters (Abell-2744 and MACSJ0416.1-240) and their associated parallel 
fields. Because the field-of-view of the ACS is larger than that of WFC3, the effective raw area of full multi-band
{\it HST} coverage is equivalent to four WFC3/IR pointings, or $\simeq 19$\,arcmin$^2$ 
(but see below, and Section 3 for the effective survey area for $z \simeq 9$ galaxies).
Within this area, the {\it HST} HFF imaging provides 7-band photometry, reaching
5-$\sigma$ depths of $B_{435} = 28.9$, $V_{606} = 29.0$, $i_{814} = 29.0$, 
$Y_{105} = 28.9$, $J_{125} = 28.5$, $J_{140} = 28.4$, and $H_{160} = 28.4$\,mag. These depths correspond
to 85\% of point-source flux-density, as scaled from aperture photometry measured on 
the HFF parallel fields through
0.4-arcsec diameter apertures in the ACS imaging, and apertures scaling from 0.4 -- 0.5-arcsec
diameter through the infrared (i.e. from $Y_{105}$ to $H_{160}$) which are designed to enclose 
70\% of point-source flux in each WFC3/IR band (see below and McLure et al. 2013). The typical depth values 
listed here are averaged between the Abell-2744 and  
MACSJ0416.1-240 imaging, with the former being typically 0.1--0.2 mag deeper in most wavebands
(due to the background).

Because we are particularly interested here in sources which are only detectable 
longward of $\simeq 1.1$\,$\mu$m, additional information at wavelengths 
longer than $H_{160}$ is still of value, even though the available 
ground-based $K$-band imaging and {\it Spitzer} IRAC imaging cannot reach the 
same depths as achieved with WFC3/IR. This is because even non-detections at these 
longer wavelengths can help to exclude potential low-redshift interlopers, and refine 
high-redshift photometric redshift solutions. We have therefore made our own reductions 
of the deep ($\simeq 25-30$\,hr), VLT Hawk-I $K_s$ observations of the HFF (HAWKI-FF, PI: Brammer, ESO Programme ID 092.A-0472), 
which provides $K_s$-band imaging across both the cluster and parallel fields, with a FWHM $\simeq$ 0.45\,arcsec,
reaching a 5-$\sigma$ detection depth of $K_s = 25.6$\,mag within a 1-arcsec diameter aperture,
(which encloses $\simeq 75$\% of point-source flux).
Although the coverage is much less extensive,
we also considered the Gemini AO imaging of the inner regions of the HFF reported by Schirmer et al. (2014), 
which has now been completed (5.2 hr of exposure) for the MACSJ0416.1-240 cluster, and reaches 
point-source depths of $K_s \simeq 25.5$\,mag.

Finally we have also mosaiced the available {\it Spitzer} IRAC 3.6\,$\mu$m and 4.5\,$\mu$m 
imaging of the HFF cluster and parallel fields. This provides imaging reaching a 5-$\sigma$ point source 
detection depth of $\simeq 26$\,mag, subject to the effectiveness of local source deconfusion.  Given the difficulties of accurately deconfusing the {\it Spitzer} data, particularly in the two cluster fields, it was not included in the final SED fits, although it was utilized as a sanity check to guard against contamination from dusty low-redshift interlopers.

\subsection{Catalogue creation}

Initial photometric catalogues were produced by running SE{\sc xtractor} 2.8.6 (Bertin \& Arnouts 1996) 
in dual-image mode, using each of the WFC3/IR filters sampling long-ward of the Lyman-break at $z\geq 8.5$
(i.e. $J_{125}, J_{140}\, \&\, H_{160}$) as the detection image. In
addition, further catalogues were produced using all of 
the relevant stacks of the {\it HST} near-infrared data as the detection image
(i.e. $J_{125}+J_{140}+H_{160}, J_{125}+J_{140} \,\&\, J_{140}+H_{160}$).

From this initial set of six object catalogues, a master catalogue
was constructed containing every unique object which was detected
at $\ge$ 5-$\sigma$ significance in any of the detection images. For those 
objects which were present in multiple catalogues, the positional 
information and photometry based on the highest signal-to-noise ratio
detection was propagated to the master catalogue. We note that all
of the final objects selected in this paper are detected at 
better than 5-$\sigma$ in the $J_{140} + H_{160}$ stack, and all 
but two are detected at better that 5-$\sigma$ in the $H_{160}$
imaging alone.

A master photometry catalogue was then constructed of all unique
objects, detected at $\geq 5\sigma$ in one of the six available
catalogues, and undetected (at $\leq 2\sigma$) in all of the ACS
$B_{435}$, $V_{606}$ \& $i_{814}$ images.

\begin{figure*}
\includegraphics[width=18.5cm, angle=0]{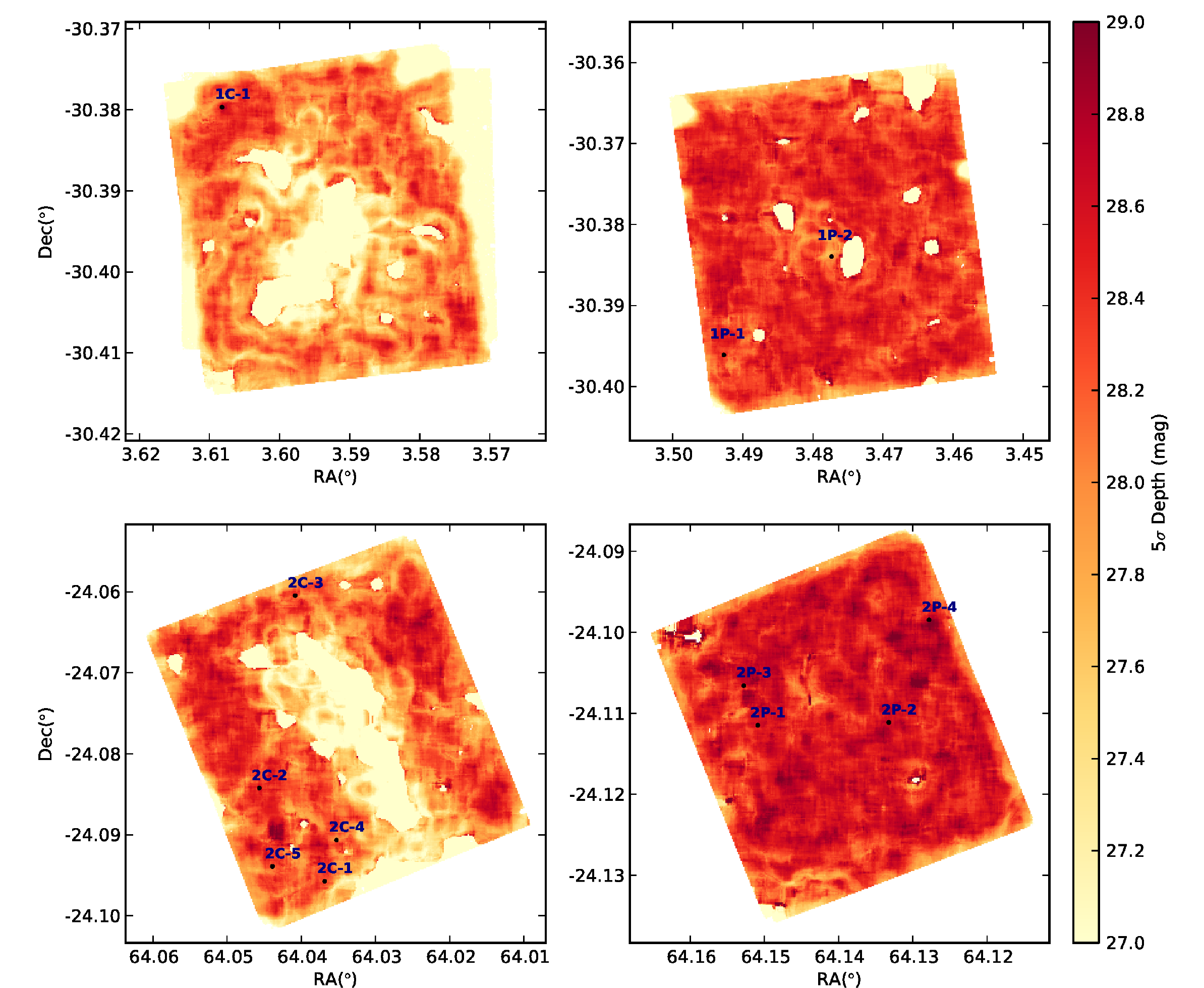}
\caption{The locations of the 12 $z \simeq 9$ galaxies uncovered in this study from the four HFF pointings
completed to date, with the positions of the objects superimposed on the depth maps of the $H_{160}$ imaging.
The colour coding indicates the 5-$\sigma$ detection depth corresponding to 85\% of point-source flux density.
As described in the text, these `local' depth measurements were established from the rms obtained from a grid of 
$\simeq 150$ nearby apertures after exclusion of any pixels containing significant source flux density.
The effective area of our survey corresponds to the sum of the regions which have depths of  28 mag or deeper
in this format, although the impact of gravitational lensing in the two cluster fields also further reduces the effective final area 
of our survey at $z \simeq 9$ (see Section 3).}
\end{figure*}

\subsection{Photometry}
For consistency with the HUDF12 analysis, the near-infrared photometry adopted in this study
is based on circular apertures, with the aperture diameter 
in each of the near-infrared bands tuned to enclose 
$\ge 70$\% per cent of the flux-density of the
filter-specific point spread function (PSF). With {\it HST} WFC3/IR 
the required aperture 
diameters are 0.40, 0.44, 0.47 and 0.50\,arcsec 
for the $Y_{105}$, $J_{125}$, $J_{140}$, and $H_{160}$ images respectively.
For the photometry in the ACS imaging we simply adopted an aperture diameter of 
0.4\,arcsec for all bands, to avoid continuing to shrink the 70\%-diameter apertures
to the point where astrometric errors become potentially problematic 
for accurate photometry. 

For the spectral energy distribution (SED) fitting we then corrected all the above aperture photometry to 85\%\ of 
total point-source flux-density using the appropriate PSF for each band.
Since the near-infrared photometric apertures have all been tuned to incorporate 
70\%\ of point-source flux-density, the correction 
to 85\%\ of flux-density corresponds to a shift of $-0.21$\,mag. In the ACS bands this 
correction obviously becomes progressively larger as we move from $i_{814}$ to $B_{435}$. In order to account for any extended flux, 
the total magnitudes reported in Table 1, and adopted for all subsequent luminosity function analysis, were calculated using  a 
large (0.85-arcsec diameter) circular aperture plus an additional point-source correction.

Finally, for completeness, we also experimented with convolving all of the 
{\it HST} imaging to the same resolution as the $H_{160}$ imaging (using an appropriate kernel 
for each waveband) before then undertaking aperture photometry in all bands with an 
aperture of constant diameter (i.e. 0.5\,arcsec). The results of the SED fitting
(described below) utilising this alternative matched-PSF photometry 
were, in practice, indistinguishable from those obtained used the matched aperture 
photometry as described above.

\begin{table*}
\caption{The final $z\simeq 9$ galaxy sample.
Column one lists the candidate names utilised in this paper and columns two and 
three list the coordinates. Columns four and five give  
the best-fitting photometric redshift and the corresponding $1\sigma$
uncertainty. Column six gives the total observed $H_{160}$ magnitude, measured
using a 0.85-arcsec diameter circular aperture, plus an additional
point-source correction for extended flux. Column seven gives the source magnification (with associated uncertainty) for the galaxies in the cluster fields 
(estimated from the CATS (Richard et al. 2014) and Zitrin et al. (2014) magnification models as discussed in the text) 
and then column eight gives the resulting final apparent magnitudes after accounting for the relevant 
de-magnification of observed brightness. The inferred total absolute UV magnitude (derived using a top-hat 
filter centred on  $\lambda_{rest} \simeq 1500$\AA) 
is given in column nine, and finally column ten
provides references to recent independent discoveries of seven of our twelve $z \simeq 9$ galaxies: (1) Zheng et
al. (2014), (2) Ishigaki et al. (2015) and (3) Coe, Bradley \& Zitrin (2015).}
\begin{tabular}{lccccccccc}
\hline
Name & RA(J2000) & Dec(J2000) & $z_{phot}$ &$\Delta z$           & $H_{160}$         & Mag              &  De-mag $H_{160}$ & $M_{1500}$ &References\\
\hline
HFF1C-9-1 & 00:14:24.93 & $-$30:22:56.14 & 8.4 & 7.9$-$8.7  & 26.75$^{+0.08}_{-0.08}$& 1.5$^{+0.6}_{-0.1}$ & 27.2$^{+0.1}_{-0.1}$& $-$19.9   &1, 2, 3 \\[1ex]
HFF1P-9-1 & 00:13:57.33 & $-$30:23:46.24 & 9.3 & 8.4$-$9.6   & 27.91$^{+0.17}_{-0.15}$& -                & 27.9$^{+0.2}_{-0.2}$& $-$19.6   &2           \\[1ex]
HFF1P-9-2 & 00:13:53.65 & $-$30:23:02.49 & 8.9 & 2.0$-$10.0 & 28.04$^{+0.24}_{-0.20}$& -                & 28.0$^{+0.2}_{-0.2}$& $-$19.2   &2          \\[1ex]
HFF2C-9-1 & 04:16:09.40 & $-$24:05:35.46 & 8.6 & 8.4$-$8.7   & 26.01$^{+0.05}_{-0.05}$& 1.7$^{+0.1}_{-0.3}$ & 26.6$^{+0.1}_{-0.1}$& $-$20.6   &\\[1ex]
HFF2C-9-2 & 04:16:11.52 & $-$24:04:54.01 & 8.5 & 8.4$-$8.6   & 26.55$^{+0.04}_{-0.04}$& 1.7$^{+0.2}_{-0.1}$ & 27.1$^{+0.1}_{-0.1}$& $-$20.2   &3 \\[1ex]
HFF2C-9-3 & 04:16:10.35 & $-$24:03:28.47 & 8.7 & 7.6$-$9.9  & 27.82$^{+0.17}_{-0.15}$& 3.3$^{+0.1}_{-0.3}$ & 29.1$^{+0.2}_{-0.2}$& $-$18.1   &\\[1ex]
HFF2C-9-4 & 04:16:09.02 & $-$24:05:17.21 & 8.5 & 8.3$-$8.7 & 27.92$^{+0.27}_{-0.21}$& 1.9$^{+0.5}_{-0.3}$ & 28.6$^{+0.3}_{-0.2}$& $-$18.7   &\\[1ex]
HFF2C-9-5 & 04:16:11.09 & $-$24:05:28.84 & 8.6 & 1.8$-$8.9 & 28.03$^{+0.28}_{-0.23}$& 1.4$^{+0.2}_{-0.1}$ & 28.4$^{+0.3}_{-0.2}$& $-$18.9   &\\[1ex]
HFF2P-9-1 & 04:16:35.97 & $-$24:06:48.06 & 8.7 & 8.3$-$9.4  & 27.69$^{+0.11}_{-0.10}$& -                & 27.7$^{+0.1}_{-0.1}$& $-$19.6   &3 \\[1ex]
HFF2P-9-2 & 04:16:31.72 & $-$24:06:46.78 & 9.0 & 8.5$-$9.5  & 28.37$^{+0.20}_{-0.17}$& -                & 28.4$^{+0.2}_{-0.2}$& $-$19.1   &\\[1ex]
HFF2P-9-3 & 04:16:36.42 & $-$24:06:30.41 & 9.0 & 8.7$-$9.5  & 28.11$^{+0.17}_{-0.15}$& -                & 28.1$^{+0.2}_{-0.2}$& $-$19.3   &3 \\[1ex]
HFF2P-9-4 & 04:16:30.43 & $-$24:06:01.16 & 8.4 & 8.0$-$8.6  & 28.08$^{+0.14}_{-0.12}$& -                & 28.1$^{+0.1}_{-0.1}$& $-$19.1   &3 \\[1ex] 
\hline\hline\end{tabular}
 \end{table*}

\subsection{Depth analysis and photometric uncertainties}
A key element in producing robust samples of high-redshift galaxy
candidates is the derivation of accurate photometric
depth measurements, both for the reliable exclusion 
of low-redshift contaminants, and to produce the robust photometric error
estimates required for accurate photometric redshifts to be derived. Following our previous work (e.g. McLure et
al. 2010, 2011, 2013) we derived accurate measurements of the depths of
each image by measuring the aperture--to--aperture rms from a grid
of apertures placed in blank-sky regions (i.e. after excluding all regions 
of the images containing significant object flux-density, as defined 
from a dilated SE{\sc xtractor} segmentation map). This technique accurately
accounts for the additional noise introduced by a variety of
systematic effects which are not captured by a simple noise model, or
even by the rms maps produced by the {\sc drizzle} data reduction
algorithm (e.g. Koekemoer et al. 2013). 

To establish the appropriate `local' aperture--to--aperture rms measurement for each object in 
each image we based the statistics on a sub-grid of the nearest $\simeq 150$ apertures.  
For consistent use within  the SED fitting this raw aperture-to-aperture rms was then scaled to 
the equivalent value for 85\%\, of PSF flux density. 

In Fig.\,1 we show the $H_{160}$ depth maps for the four fields which result from this process
of local noise estimation. As can be seen from this figure, all of the final $z \simeq 9$ candidates listed in Table 
1 and discussed in detail in Section 3 are selected from the regions of the image for which 
$\sigma_{160} > 30$\,mag within our adopted 0.5-arcsec diameter photometric aperture. 
The effective survey areas which result from confining attention to 
these deep regions of the {\it HST} imaging are summarized below in Section 3.

\subsection{Photometric redshifts}
 
After initial exclusion of obvious low-redshift contaminants, 
the photometry for our remaining potential high-redshift galaxy candidates 
was processed using the photometric redshift code described in 
McLure et al. (2011), and utilized in the HUDF12 luminosity 
function analysis undertaken by McLure et al. (2013). This code 
is described in detail in McLure et al. (2011), but its key elements 
can be summarized briefly as follows.

The multi-frequency photometry (corrected to 85\%\ of a point-source as described above) for 
each galaxy is fitted with a range of galaxy templates, either 
empirical spectra or evolving synthetic galaxy-evolution models, with
the best-fitting galaxy parameters determined via $\chi^2$ minimization.
To ensure a proper treatment of the photometric uncertainties, the
fitting is performed in flux-density--wavelength space, rather than
magnitude--wavelength space, and any measured negative flux densities
are consistently included. The code allows a wide range of 
dust reddening, based on the Calzetti et al. (2000) dust attenuation law, 
and implements IGM absorption according to the prescription of 
Madau (1995). 

In addition, the code also includes the contribution from the strongest emission lines in the rest-frame optical (i.e. [O{\sc ii}], [O{\sc iii}], H$\beta$ and H$\alpha$).  Initially, the 
H$\alpha$ line flux is calculated from the star-formation rate of the stellar population template using the Kennicutt \& Evans (2012) calibration, and the H$\beta$ line flux is then 
calculated assuming case B recombination. The strength of the [O{\sc ii}] and [O{\sc iii}] emission lines are then set by the line ratios observed in star-forming galaxies
at $z\simeq 2$ in the recent study of {\it HST} grism spectra by Cullen et al. (2014). This is particularly relevant to the current study because many of the $z\simeq 9$ candidates 
have secondary photometric redshift solutions corresponding to young, star-forming galaxies at $z\simeq 2$. If necessary, the code can also include Ly-$\alpha$ emission 
within a plausible range of rest-frame equivalent widths (chosen to be $EW_0 \leq 240$\AA; 
Charlot \& Fall 1993). The best-fitting photometric redshifts and redshift 
probability density functions adopted here are based on Bruzual \& 
Charlot (2003) stellar  population models with metallicities of 0.2$Z_{\odot}$ or $Z_{\odot}$  (although 
generally the particular choice of best-fitting 
stellar population model has little impact on the derived photometric redshift 
constraints). 

Based on the photometric redshift results, objects were excluded
if it was impossible to obtain a statistically acceptable solution at 
$z > 8$ or if the photometric redshift probability
density function indicated a very low probability that the object lies
at $z > 8$.

At this point the remaining high-redshift candidates were carefully visually inspected to exclude
any object for which even a marginal detection was apparent in a stack of all of the ACS optical 
imaging (i.e. $B_{435} + V_{606} + i_{414}$). Lastly, the 
final galaxy sample was refined down to include only those objects with a best-fitting
photometric redshift solution in the range $8.4 < z < 9.5$.

As a final check we repeated the photometric redshift analysis using the public code Le Phare\footnote{www.cfht.hawaii.edu/~arnouts/LEPHARE/lephare.html}. 
This code produced photometric redshift solutions essentially identical to those produced by our own customized code as described above, with the values 
of $z_{phot}$ produced for potential $z > 8$ galaxies by the two codes in agreement to within
$\delta z <  0.05$.

In the next section we describe the properties of the final sample of $z \simeq 9$ galaxies 
in the four HFF pointings which was produced by this process.

\begin{figure*}
\includegraphics[width=16.6cm, angle=0]{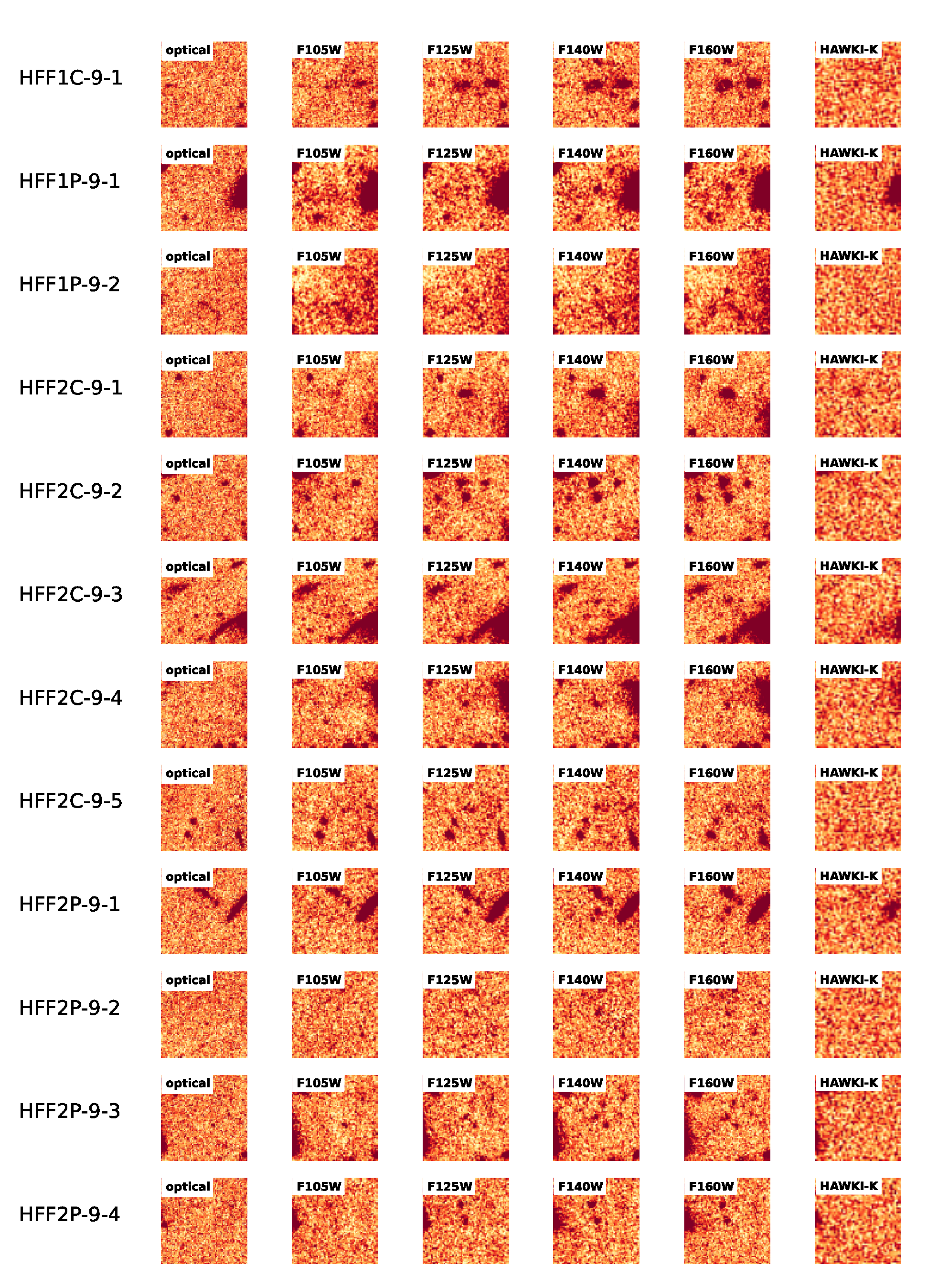}
\caption{Postage-stamp images of our 12 selected $z \simeq 9$ galaxies as listed in Table 1. Left to right we show,
for each object, an image constructed from the stacked optical bands ($B_{435}+V_{606}+i_{814}$), followed
by the four {\it HST} WFC3/IR bands, and finally the ground-based Hawk-I $K_S$ band. All images are 5 $\times$ 5 arcsec, with 
North to the top and East to the left. The colour scale was set to span $\pm$2-$\sigma$ on either side of the 
median background after removal of any significant companion-source flux in the vicinity of each high-redshift candidate.}
\end{figure*}

\noindent
\section{RESULTS}

\subsection{Final $\bmath{z \simeq 9}$ galaxy sample and survey area}

In Table 1 we list the 12 galaxies with $H_{160} < 28.6$ and robust 
photometric redshift solutions in the range $8.4 < z < 9.5$ 
uncovered from within the four completed HFF pointings. We allow our adopted redshift range for $z \simeq 9$ galaxies
to extend to $z \simeq 8.4$ because, for objects so close to the $z = 8.5$ boundary, inclusion of even 
a relatively small amount of Lyman-$\alpha$ emission in the SED templates used to derive 
the photometric redshifts would move them to $z_{phot} \ge 8.5$ (obviously, exclusion 
of the two objects with $z_{phot} \simeq 8.4$ simply reduces our $z \simeq 9$ sample from 12 to 10 galaxies, while also 
reducing slightly the effective search volume).

\begin{figure*}
\includegraphics[width=16.0cm, angle=0]{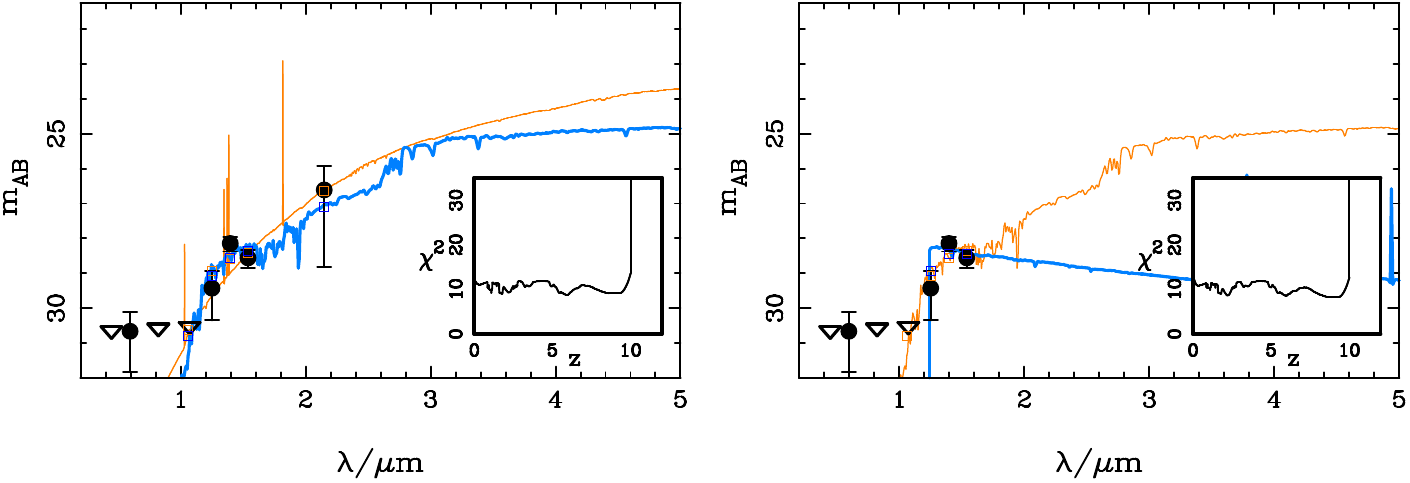}
\caption{A demonstration of why we reject one of the $z \simeq 9$ galaxies reported by Ishigaki
et al. (2015) in the Abell-2744 parallel field (their object YJ3).
On the left we show, in blue, the best-fitting $z \simeq 6$ SED, and, in orange, the alternative lower redshift ($z \simeq 2$) fit, 
which result from including our Hawk-I $K_S$-band photometry as well as the {\it HST} data, while on the right we show the corresponding fits including only the {\it HST} information.
In fact, in both cases it can be seen from the inserts of $\chi^2$ versus redshift that the redshift solutions are
poorly constrained, and so this is not in any case an attractive object for inclusion in a robust sample of objects.
However, this figure shows that even the marginal detection in the $K_S$-band tips the balance in favour of a solution at $z \simeq 6$
rather than the $z \simeq 9$ solution which is (just) favoured without the $K_S$-band information.}
\end{figure*}

The positions of these objects within the {\it HST} images are shown in Fig.\,1, where they are superimposed
on the $H_{160}$ depth maps discussed above in Section 2. For the objects selected within the cluster fields, 
we also provide estimates of source magnification, and hence de-magnified $H_{160}$ magnitudes in Table 1. To derive 
these we have utilised the magnification map at $z \simeq 9$ produced by the CATS team\footnote{http://archive.stsci.edu/prepds/frontier/lensmodels}
(e.g. Richard et al. 2014) 
and have considered the alternative Zitrin et al. (2015) 
LTM and NFW magnification maps to derive an estimate of 
the uncertainty in the source magnifications. The values given in Table 1 
are the median of these three estimates at the position of 
each object (in practice averaged across the appropriate photometric 
aperture) with the tabulated uncertainty indicating the extremes 
provided by the two alternative estimates. 

From Table 1 it can be seen that, partly because of 
our decision to confine our attention to the cleaner, consistently-deep areas of imaging, none of these objects 
has a particularly high estimated magnification; only one has a magnification greater than 2, and has 
a de-magnified $H_{160}$ magnitude fainter than $H_{160} = 28.6$. This simplifies our subsequent analysis, because 
this means that we have in practice uncovered 11 galaxies at $z \simeq 9$ with intrinsic $H_{160} < 28.6$ from a survey 
which is in effect complete to this depth (as we achieve this depth 
at better than 5-$\sigma$ in the $J_{140} + H_{160}$ stack 
across all of our selected area 
even in regions of the images where there is no magnification from gravitational lensing).  

We have also used the aforementioned CATS magnification maps to estimate the effective area of our survey
at $z \simeq 9$, and hence the inferred cosmological volume probed over the redshift range $8.4 < z < 9.5$.
Within the Abell-2744 cluster field we have uncovered one $z \simeq 9$ galaxy from a survey area of 1.85\,arcmin$^2$
(the area of the imaging which, as indicated in Fig.\,1, has a local depth deeper than $\sigma_{160} = 30$\,mag within a 
0.5-arcsec diameter aperture), which reduces
to 0.71\,arcmin$^2$ after accounting for lensing. Within the MACSJ0416.1-240 cluster field we have uncovered five 
$z \simeq 9$ galaxies (or four if attention is confined to intrinsic $H_{160}~<~28.6$\,mag) 
from a survey area of 2.59\,arcmin$^2$, which reduces to 1.09\,arcmin$^2$ after accounting for lensing. Within the Abell-2744 parallel
field we have uncovered two $z \simeq 9$ galaxies within a survey area of 4.49\,arcmin$^2$, while within the 
MACSJ0416.1-240 parallel field we have uncovered four $z \simeq 9$ galaxies within a survey area of 4.64\,arcmin$^2$.

In summary, therefore, accounting for the reduction in effective survey area due to enhanced noise, bright-object contamination,
and gravitational lensing, our final sample of $z \simeq 9$ galaxies is selected from a total survey area of 10.9\,arcmin$^2$, 
and contains 12 galaxies with observed $H_{160} < 28.6$, 11 of which have instrinsic (de-lensed) $H_{160} < 28.6$.

In Fig.\,2 we provide multi-band postage-stamp images of all 12 of our selected objects, and in Appendix A 
we show plots of the best SED fits to the photometry, along with corresponding plots of $\chi^2$ versus redshift for each source.

In Section 4 we utilise this new sample, in combination with the existing fainter HUDF12 $z \simeq 9$ galaxy 
sample (selected from an effective area of $\simeq 3$\,arcmin$^2$) to revisit 
the normalization of the $z \simeq 9$ UV galaxy LF. First, however, we consider how our sample compares 
to those selected by other authors in the HFF pointings which have already be investigated (or have previously 
been studied using the shallower CLASH data).

\subsection{Abell-2744 cluster field}

We find only one galaxy in the Abell-2744 cluster field in the redshift range of interest. This galaxy is confirmed in both
Ishigaki et al. (2015) (their Y5) and Zheng et al. (2014) (their YD4).

In this field the agreement with Ishigaki et al. (2015) is very good in the sense that we recover
all three of the objects they report in the cluster at $z = 8.2 - 8.4$ (their table 3),
but two of these (their Y2 and Y3, which is also YD8 in Zheng et al. 2014) remain at $z_{phot} = 8.2$
in our photometric redshift determination, in exact agreement with Ishigaki et al. (2015), 
and hence these are not included in Table 1, which we start at a minimum redshift of $z_{phot} = 8.4$
as discussed above.

The agreement with the $z \simeq 9$ objects listed in the Abell-2744 cluster field by Zheng et al. 
(2014) is more problematic. Other than confirming their YD4 
as discussed above (and documented in our Table 1), and confirming their YD8 to lie at $z_{phot} \simeq 8.2$, 
we fail to confirm that any of the remaining 8 objects in table 3 of Zheng et al. (2014)
in fact lie at $z > 8.2$. In several cases this is simply because we find the objects to be real high-redshift objects, but with $z_{phot}$
significantly lower than listed by Zheng et al. (2014). For example, we find their YD1 to lie at $z_{phot} = 6$, 
and their YD5, YD6, YD9 and YD10 to all lie at $7 < z < 8$. 
Two other objects (their YD3 and YD11) appear to lie at much lower
redshift ($z < 3$), while YD7 and YD8 (which correspond to Y2 and Y3 in Ishigaki et al. 2015) just escape 
inclusion with $z _{phot} = 8.2$ as already explained above. Lastly we note that YD2, with $H_{160} \simeq 28.8$\,mag,
is too faint to be included in our sample, as a result of which its redshift solution is, in any case, highly uncertain.

\subsection{Abell-2744 parallel field}

As listed in Table 1, we report two $z \simeq 9$ galaxies in the Abell-2744 parallel field. Reassuringly,
our two secure candidates correspond to the brightest two of the three reported by Ishigaki et al. (2015), 
their YJ1 and YJ2 (no analysis of the parallel field imaging was attempted by Zheng et al. 2014). 
For the third candidate reported by Ishigaki et al. (2015), their YJ3, we now find $z_{phot} \simeq 5.9$, and so
we exclude it from the list in Table 1. This object provides an example of how the inclusion of the
deep Hawk-I $K_s$-band data can tighten constraints on the photometric redshifts, and so we show in Fig.\,3
the best fitting SED (and $\chi^2$ versus redshift) for this object with and without inclusion of the
$K_s$-band photometry.  Without the $K_s$-band data, the redshift was found to be $z \simeq 9.3$.

\subsection{MACSJ0416.1-240 cluster field}
No results from other searches for high-redshift galaxies have so far been published based on the completed 
HFF imaging of the MACSJ0416.1-240 cluster field \footnote{Coe et al. (2015) identified several bright candidates 
based on partially complete data.}. However, unlike 
Abell-2744, the MACSJ0416.1-240 cluster was one of the target fields in the CLASH Treasury programme. This field 
thus featured in the CLASH survey for galaxies at $z \simeq 6 - 8$ conducted by Bradley et al. (2014) and at 
$z \simeq 9$ reported by Bouwens et al. (2014). Here we have used the new HFF data to uncover 5 galaxies 
at $z \simeq 9$, as listed in Table 1. 

\begin{figure}
\begin{centering}
\includegraphics[width=7.5cm, angle=0]{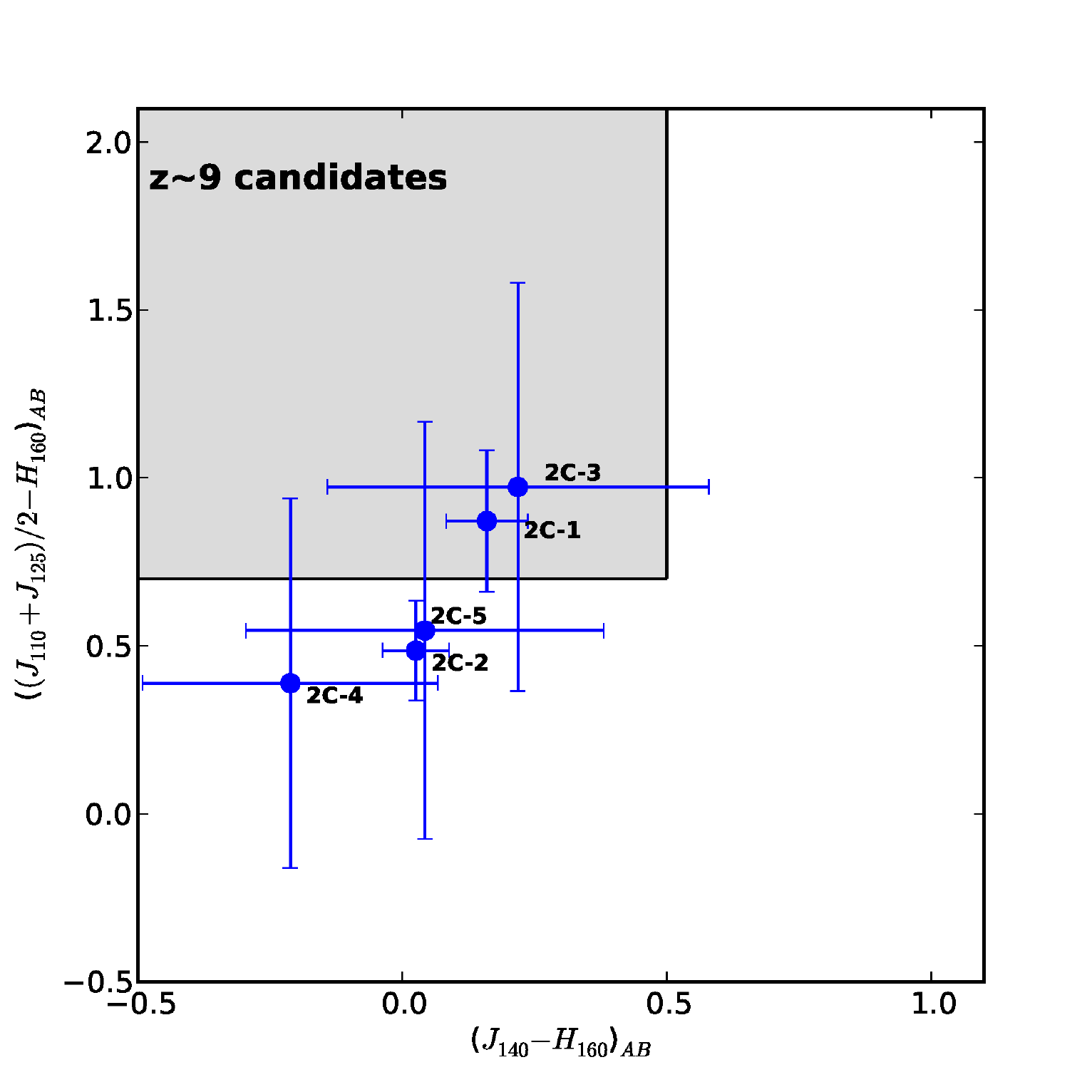}
\caption{The location of the five $z \simeq 9$ galaxies we have selected from the
MACSJ0416.1-240 cluster field relative to the colour-colour selection box used by Bouwens et al. (2014)
to select $z \simeq 9$ galaxies from the CLASH Treasury programme, which included this cluster.
The colour-colour criterion adopted by Bouwens et al. (2014) includes use of the $J_{110}$
photometry, a filter which is not used in the HFF programme, and so we are only able to show our 5 objects from this cluster here
(and this involves extracting the $J_{110}$ photometry from the CLASH data). In part because of the limited depth of
the CLASH $J_{110}$ imaging, the vertical error bars are large, but it can still be seen that
two of our objects lie within the box (we discuss in Section 3.4 why the brightest source
here would nevertheless not have been selected by Bouwens et al. 2014). The three that do not 
lie in the selection box all have photometric redshifts $z \simeq 8.5 - 8.6$, and so it 
is no surprise that they would not have been selected by this
colour-colour criterion (which is biased towards higher redshifts, 
with a selection function peaking at $z \simeq 9.2$).}
\end{centering}
\end{figure}

However, none of these were reported as $z \simeq 9$ galaxies in either of the aforementioned papers.
Since HFF2C-9-3, HFF2C-9-4 and HFF2C-9-5 all have $H_{160} \simeq 28 - 28.2$ their non-detection in the much shallower 
(1--2 orbit) CLASH imaging is not surprising. However, the two significantly brighter objects, HFF2C-9-1 and HFF2C-9-2 (which both 
have $H_{160} \simeq 26.5$) should certainly have been detected in the CLASH imaging, and so we have explored carefully
why these objects might not have been reported as $z \simeq 9$ galaxies by the CLASH studies. It transpires that 
the first of these objects, HFF2C-9-1, fell just outside the $J_{125}$ and $J_{140}$ imaging (and very close to the edge of the $H_{160}$ imaging)
obtained in the CLASH programme (although 
it is covered by the CLASH $J_{110}$ imaging), and so could not be included in the colour-colour 
criterion applied by Bouwens et al. (2014). Given that we now have coverage of this object 
in $Y_{105}$, $J_{125}$, $J_{140}$, $H_{160}$ we have combined the new HFF photometry with the CLASH 
$J_{110}$ photometry to deduce where this object would appear in the colour-selection plot utilised 
by Bouwens et al. (2014). This is shown in Fig.\,4, where we also show the other 4 $z \simeq 9$ candidates 
uncovered within this field in the current study. Fig.\,4 demonstrates that HFF2C-9-1 would indeed 
have been selected as a $z \simeq 9$ galaxy by Bouwens et al. (2014) had the complete near-infrared 
photometry been available. This plot also shows that our object HFF2C-9-3 satisfies the colour-colour
selection criterion adopted by Bouwens et al. (2014), but that our other three objects in the field (including HFF2C-9-2) do not.
However, this is not surprising, because these 3 objects lie at $z_{phot} \simeq 8.5 - 8.6$, and 
the Bouwens et al. (2014) criterion is tuned to higher redshifts (the redshift
window corresponding to the colour-colour criterion peaks at $z \simeq 9.2$).

It is therefore clear why neither HFF2C-9-1 or HFF2C-9-2 would have been selected by Bouwens et al. (2014). More confusing is why neither 
was identified as a high-redshift galaxy by Bradley et al. (2014), since both objects feature in the CLASH photometric catalogue.
In fact, in the CLASH catalogue, HFF2C-9-1 is ascribed a `low' photometric redshift of $z \simeq 2.2$, 
but it transpires that this result is driven by a claimed 6-$\sigma$ detection in the $B_{435}$ CLASH imaging. 
In fact the much deeper $B_{435}$ HFF image is completely blank (as is the full optical stack; see 
Fig.\,2), and re-inspection of the shallower $B_{435}$ CLASH 
imaging shows that if there is any flux at the object position, it appears to
be due to a linear artefact.
Given the apparently erroneous nature of the CLASH $B$-band detection, the lack of multi-band near-infrared 
photometry in the CLASH imaging, and the fact that, as stated above, this object now clearly lies within the $z \simeq 9$ 
colour-colour selection box used by Bouwens et al. (2014), we are confident that the low-redshift solution 
can be safely rejected (see Fig.\,A1). Object HFF2C-9-2 is listed in the CLASH  
catalogue as lying at $z \simeq 1.3$, but even using the CLASH photometry our photometric-redshift code yields
$z \simeq 8.45 \pm 0.35$. The new deeper HFF photometry simply further strengthens the high-redshift solution (i.e. as with object HFF2C-9-1
the much deeper ACS imaging shows no evidence for any optical detection of object HFF2C-9-2).

For completeness, we note that two other potential high-redshift galaxies in this field were ultimately removed 
in the final stages of our sample refinement. One was a nearby companion of HFF2C-9-2
situated at RA 04:16:11.53, Dec $-$24:04:53.14.  
Although this object has an acceptable and apparently robust redshift solution of $z_{phot}=8.7\pm 0.1$, 
we have reservations about its validity due to an apparent, albeit marginal ($\simeq 2\sigma$), detection in the $B_{435}$ image\footnote{Note this object was also
identified using CLASH data by Bouwens et al. (2014).}. 
The second object (at RA\,04:16:36.39 Dec\,$-$24:06:29.62) is close to HFF2P-9-3 and was 
found to have a preferred photometric redshift solution of $z_{phot}=8.9$. However, this is a very 
insecure candidate with a very large acceptable range in photometric redshift, and hence cannot be considered 
robust given the available data. We note that inclusion of these rejected objects would simply further strengthen the case
for a gradual decline in the abundance of star-forming galaxies
beyond $z\simeq 8$ (see Section 4).

\subsection{MACSJ0416.1-240 parallel field}
No previous search for high-redshift galaxies in this field has been published based on the full-depth HFF data, although Coe et al. (2015) identified several candidates 
based on partially complete data. As listed in Table 1, we report four $z \simeq 9$ galaxies.

\section{Analysis \& Discussion}

\subsection{A new estimate of the $\bmath{z \simeq 9}$ luminosity function}
We have combined the new sample of $z \simeq 9$ galaxies with the fainter sample derived through the HUDF12 programme
(Ellis et al. 2013; McLure et al. 2013) to produce a new estimate of the galaxy UV LF. We used the effective de-lensed 
area calculations described in Section 3.1 to calculate cosmological volumes for the HFF sample, and performed a new set 
of source injection and retrieval simulations with the HFF imaging to estimate completeness corrections as a function 
of magnitude (following McLure et al. 2013). Although these simulations do not include a correction for lensing shear, 
by restricting our analysis to the parallel fields and the low magnification regions of the cluster images, we can be confident 
that any shear correction is expected to be negligible (Oesch et al. 2014b).
For the fainter HUDF12 sample we simply retained the stepwise maximum likelihood results presented by McLure et al. (2013) as the best available 
estimate of the UV LF around $M_{1500} \simeq -18$.

The final results of this process are shown in Fig.\,5, where we are now able to present four bins in luminosity at $z \simeq 9$.
Clearly the two new brighter bins provided by the wider area HFF survey are consistent with the deeper HUDF12 results, and to put 
this result in context, we also show in Fig.\,5 the latest UV LFs at $z \simeq 5,6,7$ from Bowler et al. (2014a, 2014b) and $z \simeq 
8$ from McLure et al. (2013). Despite our now enlarged $z \simeq 9$ sample it is clear that, at this redshift, the data 
do not span sufficient dynamic range, and in particular do not probe sufficiently far beyond the apparent break luminosity
to undertake a meaningful new determination of the faint-end slope, $\alpha$. Therefore we proceed conservatively, and assume 
that the basic shape of the LF is unchanged since $z \simeq 8$, and fit the $z \simeq 9$ LF assuming the two extremes of
density evolution or luminosity evolution of the $z \simeq 8$ LF. These alternative fits are shown in Fig.\,5 
by the solid and dashed lines respectively, and the corresponding derived values of $M_{1500}^{\star}$ and $\phi^{\star}$
are listed in Table\,2 (in this table the errors on $\phi^{\star}$ include an 
estimated extra contribution due to cosmic variance; see Section 4.4 and Robertson et al. 2014).

\begin{table}
\caption{The Schechter-function parameters for the $z=9$ UV galaxy
luminosity function (LF) derived from the combined HFF + HUDF12 samples, and 
assuming either density evolution or luminosity evolution from the $z \simeq 8$ LF 
derived by McLure et al. (2013), with the faint-end slope locked at the $z \simeq 8$ value.
The errors on $\log(\phi^{\star})$ include the estimated impact of 
cosmic variance following Robertson (2010a,b) and Robertson et al. (2014); see section 4.4.
The units of $\phi^{\star}$ are Mpc$^{-3}$.}

\begin{tabular}{lccc}
\hline
&$M_{1500}^{\star}$ & $\log(\phi^{\star})$  & $\alpha$\\
\hline
Density Evolution & $-20.1$ & $-3.60^{+0.20}_{-0.35}$ & $-2.02$ (fixed)\\
Luminosity Evolution & $-19.7$ & $-3.35^{+0.15}_{-0.20}$ & $-2.02$ (fixed)\\
\hline
\end{tabular}
\end{table}

\begin{figure}
\includegraphics[width=8.0cm, angle=0]{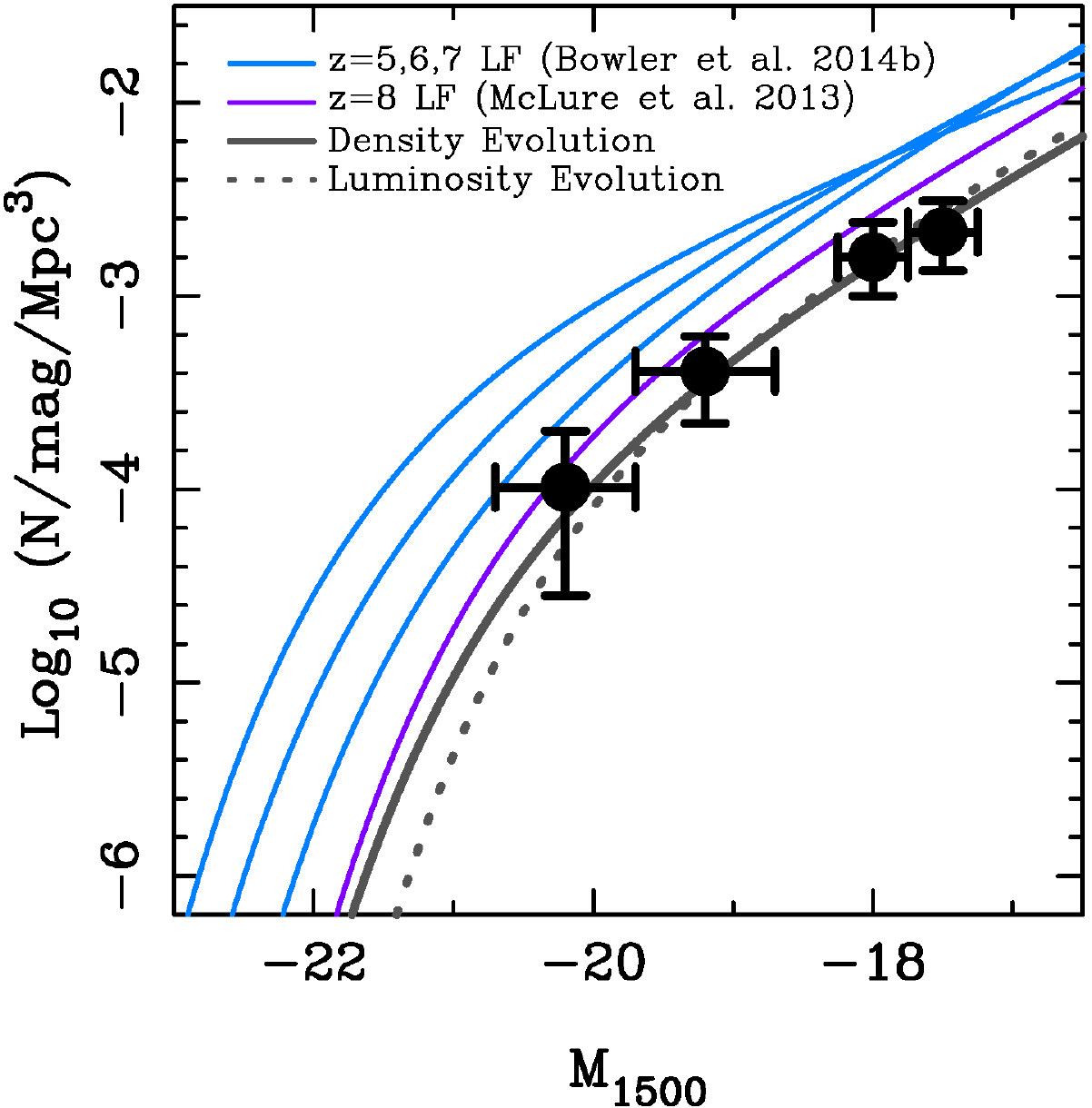}
\caption{Our new determination of the galaxy UV LF at $z \simeq 9$. The light blue curves show the LF at $z \simeq 5,6,7$ as determined by Bowler
et al. (2014a, 2014b) while the purple curve shows the UV LF at $z \simeq 8$ as determined by McLure et al. (2013). The two data-points around
$M_{1500} \simeq -18$ show our SWML estimate of the $z \simeq 9$ luminosity function as derived by McLure et al. (2013) from the HUDF12 dataset.
The two new datapoints at brighter absolute magnitudes are based on the galaxies in our new HFF sample which lie at $8.5 \le z < 9.5$ 
and have intrinsic $H_{160} < 28.6$ mag. To show the inferred LF, we lock the faint-end slope at the value determined at $z \simeq 8$, and show the effects
of assuming density (solid line) or luminosity (dotted line) evolution of the LF from $z \simeq 8$ to $z \simeq 9$.
Our new HFF points are clearly consistent with the fainter HUDF12 determination, and now marginally favour
the density-evolution fit. Integration of this function then produces a consistent but somewhat higher estimate
of UV luminosity density at $z \simeq 9$ than reported by McLure et al. (2013) (see Fig.\,6, and Section 4.2 for details).}
\end{figure}

We note that, although these two alternative fits are both acceptable, the new HFF data mean that the density evolution fit 
is slightly favoured. Continued luminosity evolution, as generally found at lower redshifts $z \simeq 5 - 8$
(e.g. Bouwens et al. 2007; McLure et al. 2009; Bouwens et al. 2011b; Bowler et al. 2014b) is, however, also still clearly  
consistent with the available data. Regardless of the precise values of the fitted Schechter-function parameters, it is clear
from Fig.\,5 that nothing too dramatic occurs between $z \simeq 8$ and $z\simeq 9$, and the galaxy UV LF continues the smooth 
decline witnessed over $z \simeq 5 - 8$. This conclusion contrasts somewhat with that reached by Oesch et al. (2013) who 
reported that the $z \simeq 9$ LF lay significantly below what is `expected' based on continued smooth evolution.
However, the sample on which Oesch et al. (2013) based this conclusion was significantly smaller than that reported here, and they did not have the 
benefit of the $J_{140}$ imaging which is so crucial in the selection of reliable $z \simeq 9$ galaxy samples. In any case, examination 
of their binned data at $z \simeq 9$ (shown in their figure 11) shows that all three of their data points 
are in fact clearly consistent with the new more accurate $z \simeq 9$ LF derived here.

\begin{figure*}
\includegraphics[width=17.0cm, angle=0]{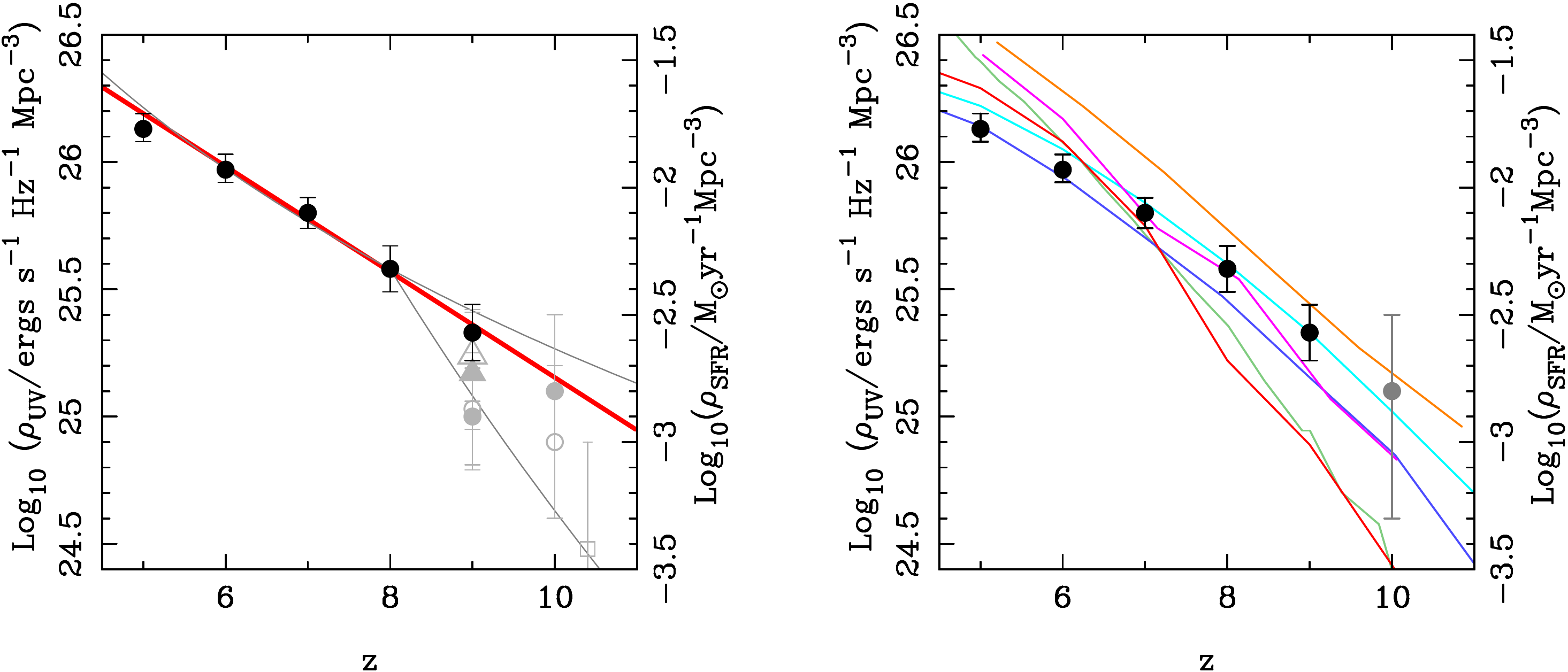}
\caption{The evolution of the observed comoving UV luminosity density (left-hand axis on each panel) and hence 
inferred star-formation rate density (right-hand axis on each panel) versus redshift (from $z \simeq 5$ to $z \simeq 10$);
$\rho_{UV}$ was converted to $\rho_{SFR}$ assuming a Salpeter IMF and the UV-to-SFR conversion of Madau, Pozzetti \& Dickinson (1998).
The filled black data-points result from the luminosity-weighted integration of the UV LFs shown in Fig.\,5 
down to $M_{1500} = -17.7$, or equivalently a limiting star-formation rate of $SFR \simeq 0.7\,{\rm M_{\odot} yr^{-1}}$. The grey points at $z \simeq 9$ and $z\simeq 10$ (open circles, filled circles, open triangle, filled triangle and open square) are taken from Ellis et al. (2013), Oesch et al. (2014b), McLure et al. (2013), Ishigaki et al. (2015) and Bouwens et al. (2015) respectively.
Our new estimate of the UV luminosity density at $z \simeq 9$ is based on the density evolution fit 
shown in Fig.\,5, with the error-bar including the effects of uncertainty in the faint-end slope, and also estimated 
cosmic variance as described in Robertson et al. (2014). The solid red line shown in the left-hand panel shows that 
a straightforward linear function of redshift (given in the text) provides an excellent description of the gradual decline in 
$\rho_{UV}$ from $z \simeq 6$ to at least $z \simeq 9$ given our new result, and apparently also to $z \simeq 10$ (Oesch et al. 2014b), although it does
not agree well with the (albeit highly uncertain) $z=10.4$ data-point from Bouwens et al. (2015). The higher and lower thin grey lines in the left-hand panel show the 
alternative extrapolations beyond $z \simeq 8$ as proposed by Oesch et al. (2014b), with $\rho_{UV}  \propto (1+z)^{-3.6}$
(higher/shallower grey line) and $\rho_{UV} \propto (1 + z)^{-10.9}$ (lower/steeper grey line); clearly our new data-point at $z \simeq 9$ favours the 
more gently declining of these extreme alternatives (although it lies slightly below this, and is better described by the aforementioned 
linear relation with $z$). In the right-hand panel we compare the observed smooth decline of UV luminosity 
density with the predictions of six recent (and very different) models of galaxy formation. The orange, cyan, pink, blue, green and red curves show the predictions from Behroozi \& Silk (2015), Cai et al. (2014), 
Khochfar et al. (in preparation), Dayal et al. (2014), Genel et al. (2014) and Henriques et al. (2014) respectively. All models have been integrated down to an effective limit of $SFR \simeq 0.7\,{\rm M_{\odot} yr^{-1}}$ and 
corrected to a Salpeter IMF where necessary. In the right-hand panel we again show our own data-points in the redshift range $5<z<9$ (black) and the latest $z\simeq 10$ estimate (grey) from Oesch et al. (2014b).}
\end{figure*}

\subsection{Inferred high-redshift decline of UV luminosity density}
Given our improved knowledge of the $z \simeq 9$ LF, it is clearly of interest to recalculate the inferred 
UV luminosity density, and hence star-formation rate density at $z \simeq 9$. As described in Robertson et al. (2010, 2013)
the high-redshift evolution of UV luminosity density has important implications for reionization, given the apparent 
need for reionization to be an extended process to explain the optical-depth to Thomson scattering inferred from the polarization of the microwave 
background (Hinshaw et al. 2013; Planck Collaboration 2014). 
It can also provide a useful and fairly straightforward test of the ability of the latest galaxy simulations to predict
the early build-up of the galaxy population.

Therefore, in Fig.\,6 we present the results of integrating the UV LFs shown in Fig.\,5 to derive the 
high-redshift evolution of UV luminosity density, $\rho_{UV}$, and hence star-formation rate density, $\rho_{SFR}$ (with the 
latter quantity derived assuming a Salpeter IMF and the UV-to-SFR conversion of Madau, Pozzetti \& Dickinson 1998). 
Our new estimated value at $z \simeq 9$ is $\log_{10} \rho_{UV} = 25.35 \pm 0.10$ 
(or, equivalently, $\log_{10} \rho_{SFR} = -2.55 \pm 0.10$) if, following what has become 
common practice, the UV LF is integrated down to $M_{1500} = -17.7$, equivalent to
$SFR \simeq 0.7\,{\rm M_{\odot} yr^{-1}}$. This limit 
to the integration facilitates comparison with previous results, but is also well motivated in the current case because, as can be seen from Fig.\,5, 
it corresponds well to the faintest absolute magnitude probed directly by the available galaxy samples at $z \simeq 9$. 

The value of $\rho_{UV}$ at $z \simeq 9$ inferred from the HUDF12 programme was already at the higher end  
of the published estimates, but the result presented in Fig.\,6 is somewhat higher still. This reflects 
the fact that, at least at the brighter luminosities, the UDF has proved to be somewhat underdense at $z \simeq 9$ 
as judged against the wider area probed by the HFF pointings studied here. Consequently, it can be seen that
our new results now strongly favour a continued smooth decline in $\rho_{UV}$ between $z \simeq 8$ and $z \simeq 9$, rather than 
a marked drop-off at this epoch. To help focus the debate over what happens to star-formation rate density beyond 
$z \simeq 8$, Oesch et al. (2013) proposed that continued smooth evolution at $z > 8$ could be parameterized as 
$\rho_{SFR} \propto (1 + z)^{-3.6}$ while their own results favoured a change at $z > 8$ to a much steeper decline, with 
$\rho_{SFR} \propto (1 + z)^{-11.4}$ (revised slightly to $\rho_{SFR} \propto (1 + z)^{-10.9}$ by Oesch et al. 2014b). 
We therefore include both these curves in the left-hand panel of Fig.\,6, to help clarify the comparison.
Our new result at $z \simeq 9$ clearly favours the shallower decline, but in fact our best fit value lies slightly 
below the $\rho_{SFR} \propto (1 + z)^{-3.6}$ extrapolation. However, it is not clear that this paramaterization provides 
the fairest or most appropriate baseline for judging whether high-redshift evolution is proceeding as `expected'. In fact, as shown
in Fig.\,6, a better description of the smooth decline of $\rho_{UV}$ from $z \simeq 6$ to $z \simeq 9$ is provided by a simple linear
relation between $\log_{10} \rho_{UV}$ and redshift, with a least-squares fit to the datapoints shown in Fig.\,6 yielding the 
solid red line shown in the left-hand panel, given by the relation $\log_{10}(\rho_{UV})= -0.208(\pm 0.035)z + 27.234(\pm 0.244)$. 
It can also be seen that this smoothly-declining relation leads naturally on to the latest estimate of $\rho_{SFR}$ at $z \simeq 10$ reported by Oesch et al. (2014b) (based on the $z \simeq 9.8$ object identified by 
Zitrin et al. 2014), although it does not agree well with the, albeit highly uncertain, estimate from Bouwens et al. (2015), plotted as the grey open square.
In addition, it can be seen from Fig. 6 that our derived star-formation rate density at $z \simeq 9$ is only 0.16 dex higher than that derived by Ishigaki et al. (2015). They choose 
to describe their result as being supportive of a steep decline in $\rho_{SFR}$ at high redshift as proposed by Oesch et al. (2013, 2014b). However, it is noteworthy that the Ishigaki et al. (2015) results are 
based only on the Abell-2744 HFF fields, which have a significantly lower number density of $z\simeq9$ galaxies than the MACSJ0416 HFF fields.

It is clear that our new $\rho_{SFR}$ determination at $z\simeq 9$, based on the largest and most
robust $z \simeq 9$ galaxy sample assembled to date, lies at the high end of pre-existing estimates. As shown by the solid red line in the 
left-hand panel of Fig. 6, our new result is in good agreement with a simple extrapolation of the evolution in $\rho_{SFR}$ seen at lower redshifts, and provides no evidence for
a dramatic decline in $\rho_{SFR}$ at $z\geq 8$. We conclude that Fig.\,6 now provides a convincing case that the smooth decline
of $\rho_{SFR}$ seen from $z \simeq 5$ to $z \simeq 8$ continues to at least $z \simeq 9$.

\subsection{Comparison with model predictions}

A detailed comparison with the full range of available theoretical models/simulations of early galaxy 
formation/evolution is beyond the scope of the present paper, but in the right-hand panel of Fig.\,6 we include a comparison 
of the observed smooth decline of UV luminosity density (and hence SFR density) with the predictions of six recent (and very different) models of galaxy formation. 
All of the different models have been integrated down to an effective limit of $SFR \simeq 0.7\,{\rm M_{\odot} yr^{-1}}$ and 
corrected to a Salpeter IMF where necessary. 

It is very clear from the right-hand panel of Fig.\,6 that, contrary to some recent reports in the literature (e.g. Oesch et al. 2014b), the Illustris model (Genel et al. 2014)
does not reproduce the observed gentle decline of UV luminosity density very well. The data and the model predictions
are in fairly good agreement at $z \simeq 6-7$ but at higher redshifts the model predictions descend too rapidly. It is interesting to note that
this disagreement is not just relative to our new measurement at $z \simeq 9$, but is already apparent by $z \simeq 8$, where 
there is a clear consensus over the observed value. The origin of this disagreement is not clear, but we note that the Illustris simulation, as is common practice, assumes 
a reionization history that implements a redshift-dependent UV background that is not self-consistent
with the computed evolving star-formation rate density. We can speculate that this internal inconsistency and/or 
the specific implementation of feedback may impact on the ability of 
the Illustris and other similar simulations to correctly predict the evolving abundance of faint galaxies at early times.

It can be seen that the observed evolution of UV luminosity density is rather better produced by the semi-analytic models of Cai et al. (2014) and Dayal et al. (2014).
It should be noted that while the Dayal et al. (2014) model has been tuned to fit the $z \simeq 7$ UV LF, it contains only two redshift independent free parameters 
designed to describe the key physics of early galaxy formation as embedded within a dark-matter halo merger tree. 
Thus it can be seen that the observed early evolution of SFR density is broadly as expected given the predicted hierarchical growth
of underlying structure within the $\Lambda CDM$ model. However, our new results suggest that our understanding of the baryonic 
physics of early galaxy growth still needs further refinement, given that globally-averaged star formation activity
appears to have built up more rapidly at very early times ($z > 10$) than predicted in current models.

\subsection{Cosmic variance and future projections}

As mentioned above, in constructing our luminosity function and UV luminosity density estimates, we
have attempted to account for cosmic variance uncertainty owing to the
luminosity-dependent spatial clustering of high-redshift galaxies. We adopt
the cosmic variance model described by Robertson (2010a,b) that utilizes
abundance-matching of galaxies and dark matter halos to estimate the
clustering bias of objects in our sample and calculate the combined
Poisson and sample variance uncertainty of the binned number counts
(see, e.g., Schenker et al. 2013). For our samples in the HFF lensing clusters,
we have accounted for the amplification of cosmic variance owing to the impact
of magnification on the effective search area following Robertson et al. (2014).
The cosmic variance uncertainty in the luminosity function shown in Fig.\,5 has
been propagated through to the uncertainty in the UV luminosity density shown
in Fig.\,6. If the Abell-2744 and MACS J0416.1-240 samples are representative
of the remaining HFFs, we project that the associated uncertainties may improve
by as much as a factor of two in the final samples.

We conclude by emphasizing the value of the HFF parallel fields in this work; these have provided the majority 
of the boost in survey area (and hence cosmological volume) for probing the brighter end of the $z \simeq 9$ LF, 
in effect more than trebling the area of usefully deep $J_{140} + H_{160}$ imaging over that previously 
provided by HUDF12. This work thus provides strong motivation for completing all six of the planned HFF 
cluster + parallel fields, and we look forward to this being achieved in the coming years.
 
\section{Conclusion}

We have completed a new search for galaxies at redshift
$z \simeq 9$ in the first two Hubble Frontier Fields for which the {\it HST} WFC3/IR and ACS imaging has now been completed
(in both the cluster and parallel fields). To ensure robust photometric redshift solutions
and to minimize incompleteness, we have confined our search to objects with $H_{160} < 28.6$\,AB mag, and have 
considered only the homogeneously deep regions of the imaging which 
have an rms noise $\sigma_{160} > 30$\,mag (within a 0.5-arcsec diameter aperture). We have also 
insisted on detections in both $H_{160}$ and $J_{140}$, to ensure the selection of 
$z \simeq 9$ objects with a confirmed blue continuum longward of the 
putative redshifted Lyman break.

The result is a survey covering an effective area (after accounting for magnification) of 10.9\,arcmin$^2$,
which yields 12 galaxies at $8.4 < z < 9.5$. Within the Abell-2744
cluster and parallel fields we confirmed the three brightest objects reported by Ishigaki et al. (2015),
but have recovered only one of the four $z > 8.4$ sources reported by Zheng et al. (2014).
In the MACSJ0416.1-240 cluster field we have uncovered five objects, and explain why each of these eluded detection
or classification as $z \simeq 9$ galaxies in the published searches of the shallower CLASH data.
Finally we have discovered four $z \simeq 9$ galaxies from within the previously unsearched
MACSJ0416.1-240 parallel field. 

Based on the published magnification maps we have deduced that only one of these 12 galaxies is 
likely boosted by more than a factor of two by gravitational lensing.
Consequently we have been able to perform a fairly straightforward reanalysis of the normalization of the $z \simeq 9$ UV galaxy
luminosity function as explored previously in the HUDF12 programme, and have then integrated the UV LF to derive the most accurate measurement 
to date of UV luminosity density (and hence star-formation rate density) at $z \simeq 9$. 

Our main conclusion is that the new data strengthen the evidence for a continued smooth decline in UV luminosity density (and hence
star-formation rate density) from $z \simeq 8$ to $z \simeq 9$, contrary to recent reports of a marked drop-off at these redshifts.
This result alleviates concerns that such a rapid down-turn in early star formation would present 
significant obstacles for the conclusion that high-redshift galaxies provided the bulk of the ultraviolet 
photons required to produce and maintain cosmic reionization.     

Finally we find that the observed decline of star-formation rate density with increasing redshift over $z \simeq 5 - 9$ is significantly
more gradual than predicted by some of the latest theoretical simulations of galaxy formation (e.g. Genel et al. 2014; Henriques et al. 2014). 
Whatever the origin of this discrepancy, our results indicate that galaxy evolution, and hence globally-averaged star formation activity
builds up more rapidly at very early times ($z > 10$) that predicted in current models.

\section*{acknowledgements}
DJM and RJM acknowledge the support of the European Research Council via the award of a 
Consolidator Grant (PI McLure). JSD and TAT acknowledge the support of the European Research Council via the award of an 
Advanced Grant (PI Dunlop). JSD also acknowledges the contribution of the EC FP7 SPACE project ASTRODEEP (Ref.No: 312725). 
B.E.R. is supported in part by the National Science Foundation under grant
No. 1228509,  and by the Space Telescope Science Institute under award
HST-GO-12498.01-A.
We thank Shy Genel and Mark Vogelsberger for providing the predictions for the Illustris simulation included in Fig.\,6.
This work is based in part on observations made with the NASA/ESA {\it Hubble Space Telescope}, which is operated by the Association 
of Universities for Research in Astronomy, Inc, under NASA contract NAS5-26555.
This work is also based in part on observations made with the {\it Spitzer Space Telescope}, which is operated by the Jet Propulsion Laboratory, 
California Institute of Technology under NASA contract 1407. This work uses data taken with the Hawk-I instrument on the European Southern 
Observatory's Very Large telescope from ESO programme: 092.A-0472.

{}

\appendix

\section{Spectral energy distribution fits and photometric redshifts}

In this Appendix we show, for completeness, the spectral energy distribution (SED) fits 
to the multi-wavelength photometry of the twelve objects listed in Table 1, to illustrate the 
photometric redshift constraints which led to their classification as likely galaxies in the 
redshift range $8.4 < z < 9.5$. From Fig. A1 it can be seen that most of the objects have secure 
$z \simeq 9$ solutions, although three of the faintest sources also have acceptable lower redshift solutions.

\begin{figure*}
\includegraphics[width=17.5cm, angle=0]{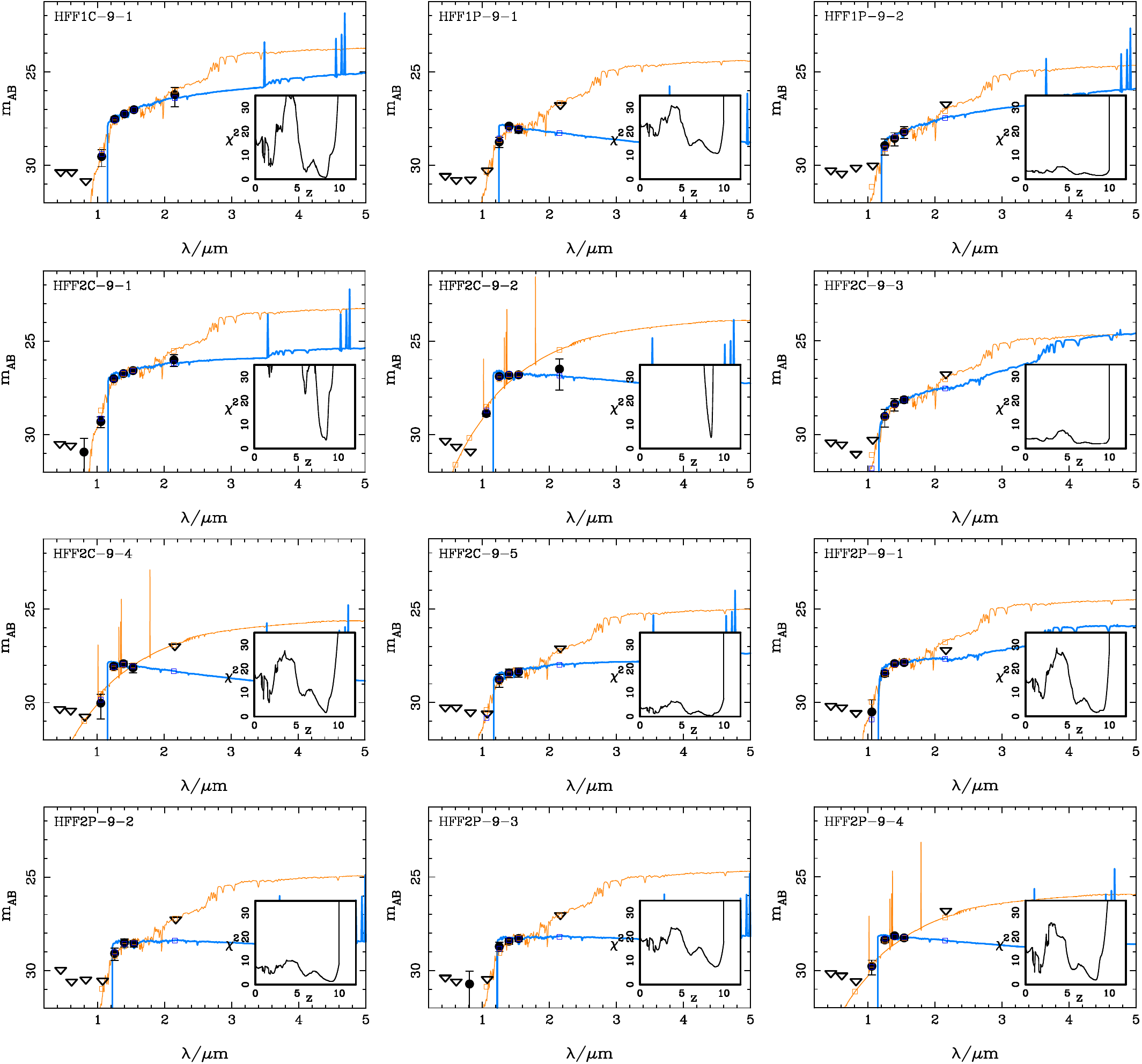}
\caption{The spectral energy distribution (SED) fits to the multi-waveband photometry for the 
twelve $z \simeq 9$ galaxies selected in this study as listed in Table 1. In each panel we 
show the preferred high-redshift SED in blue, and the alternative lower redshift solution in orange. Arrow heads 
indicate the (local) 1-$\sigma$ upper limits derived from non-detections in the relevant photometric bands.
The insert in each plot shows $\chi^2$ versus redshift. In general, it can be seen that the alternative lower redshift solutions
lie either at $z \simeq 2$ or $z \simeq 6$. However, we note that the physical plausibility of several of these 
solutions is debatable,
with the $z \simeq 6$ solutions often corresponding to maximally old quiescent stellar populations, while 
several of the $z \simeq 2$ solutions invoke extreme reddening combined with strong line emission.} 
\end{figure*}


\begin{thebibliography}{} 
\bibitem{ } Atek H., et al., 2015, AJ, 800, 18
\bibitem{ } Behroozi P., Silk J., ApJ, 799, 32
\bibitem{ } Bertin E., Arnouts S., 1996, A\&AS, 117, 393
\bibitem{ } Bouwens R.J., Illingworth G.D., Franx M., Ford H., 2007, ApJ, 662, 72 
\bibitem{ } Bouwens R.J., et al., 2010, ApJ, 709, L133
\bibitem{ } Bouwens R.J., et al., 2011a, Nat, 469, 504
\bibitem{ } Bouwens R.J., et al., 2011b, ApJ, 737, 90 
\bibitem{ } Bouwens R.J., et al., 2014, ApJ, 795, 126
\bibitem{ } Bouwens R.J., et al., 2015, ApJ, 803, 34
\bibitem{ } Bowler R.A.A., et al., 2012, MNRAS, 426, 2772
\bibitem{ } Bowler R.A.A., et al., 2014, MNRAS, 440, 2810
\bibitem{ } Bowler R.A.A., et al., 2014, arXiv:1411.2976
\bibitem{ } Bradley L.D., et al., 2012, ApJ, 760, 108
\bibitem{ } Bradley L.D., et al., 2014, ApJ, 792, 76
\bibitem{ } Bruzual G., Charlot S., 2003, MNRAS, 344, 1000
\bibitem{ } Bunker A.J., et al., 2010, MNRAS, 409, 855
\bibitem{ } Cai Z.-Y., et al., 2014, ApJ, 785, 65
\bibitem{ } Calzetti D., Armus L., Bohlin R.C., Kinney A.L., Koornneef J., Storchi-Bergmann T., 2000, ApJ, 533, 682
\bibitem{ } Charlot S., Fall M.S., 1993, ApJ, 415, 580
\bibitem{ } Coe D., et al., 2013, ApJ, 762, 32 
\bibitem{ } Coe D., Bradley L., Zitrin A., 2015, ApJ, 800, 84
\bibitem{ } Cullen F., Cirasuolo M., McLure R.J., Dunlop J.S., Bowler R.A.A., 2014, MNRAS, 440, 2300 
\bibitem{ } Dayal P., Ferrara A., Dunlop J.S., Pacucci F., 2014, MNRAS, 445, 2545
\bibitem{ } Dunlop J.S., 2013, ASSL, 396, 223
\bibitem{ } Dunlop J.S., McLure R.J., Robertson B.E., Ellis R.S., Stark D.P., Cirasuolo M., de Ravel L., 2012, MNRAS, 420, 901
\bibitem{ } Dunlop J.S., et al., 2013, MNRAS, 432, 3520
\bibitem{ } Ellis R.S., et al., 2013, ApJ, 763, L7
\bibitem{ } Finkelstein S.L., et al., 2010, ApJ, 719, 1250
\bibitem{ } Finkelstein S.L., et al., 2014, arXiv:1410.5439
\bibitem{ } Genel S., et al., 2014, MNRAS, 445, 175
\bibitem{ } Grogin N.A., et al., 2011, ApJS, 197, 35
\bibitem{ } Henriques B., et al., 2014, arXiv:1410.0365
\bibitem{ } Hinshaw G., et al., 2013, ApJS, 208, 19
\bibitem{ } Ishigaki M., et al., 2015, ApJ 799, 12
\bibitem{ } Jaacks J., Choi J.-H., Nagamine K., Thompson R., Varghese S., 2012, MNRAS, 420, 1606
\bibitem{ } Kimm T., Cen R., 2013, ApJ, 776, 35
\bibitem{ } Kennicutt R.C., Evans N.J.,  2012, ARA\&A, 50, 531
\bibitem{ } Koekemoer A.M., et al., 2011, ApJS, 197, 36
\bibitem{ } Koekemoer A.M., et al., 2013, ApJS, 209, 3
\bibitem{ } Madau P., 1995, ApJ, 441, 18
\bibitem{ } Madau P., Pozzetti L., Dickinson M., 1998, ApJ, 498, 106
\bibitem{ } McLure R.J., Cirasuolo M., Dunlop J.S., Foucaud S., Almaini O., 2009, MNRAS, 395, 2196 
\bibitem{ } McLure R.J., Dunlop J.S., Cirasuolo M., Koekemoer A.M., Sabbi E., Stark D., Targett T.A., Ellis R.S., 2010, MNRAS, 403, 960 
\bibitem{ } McLure R.J., et al., 2011, MNRAS, 418, 2074
\bibitem{ } McLure R.J., et al., 2013, MNRAS, 432, 2696
\bibitem{ } Oesch P.A., et al., 2010, ApJ, 709, L16
\bibitem{ } Oesch P.A., et al., 2012a, ApJ, 745, 110
\bibitem{ } Oesch P.A., et al., 2012b, ApJ, 759, 135
\bibitem{ } Oesch P.A., et al., 2013, ApJ, 773, 75
\bibitem{ } Oesch P.A., et al., 2014a, ApJ, 786, 108
\bibitem{ } Oesch P.A., et al., 2014b, arXiv:1409.1228
\bibitem{ } Oke J.B., 1974, ApJS, 27, 21
\bibitem{ } Oke J.B., Gunn J.E., 1983, ApJ, 266, 713
\bibitem{ } Paardekooper J.-P., Khochfar S., Dalla Vacchia C., 2013, MNRAS, 429 L94
\bibitem{ } Planck Collaboration, Ade P.A.R., et al., 2014, A\&A, 571, 15 
\bibitem{ } Richard J., et al. 2014, MNRAS, 444, 268
\bibitem{ } Robertson B., Ellis R.S., Dunlop J.S., McLure R.J., Stark D., 2010, Nat, 468, 49
\bibitem{ } Robertson B., 2010a, ApJ, 713, 1266
\bibitem{ } Robertson B., 2010b, ApJL, 716, L229
\bibitem{ } Robertson B., et al., 2013, ApJ, 768, 71
\bibitem{ } Robertson B., et al., 2014, ApJL, 796, L27
\bibitem{ } Schenker M.A., et al., 2013, ApJ, 768, 196
\bibitem{ } Schirmer M., et al., 2014, arXiv:1409.1820
\bibitem{ } Schmidt K.B., et al., 2014, ApJ, 786, 57
\bibitem{ } Trenti M., et al., 2011, ApJ, 727, L39
\bibitem{ } Vogelsberger M., et al., 2014, MNRAS, 444, 1518 
\bibitem{ } Zheng W., et al., 2012, Nat, 489, 406
\bibitem{ } Zheng W., et al., 2014, ApJ, 795, 93
\bibitem{ } Zitrin A., et al., 2014, ApJ, 793, L12
\bibitem{ } Zitrin A., et al., 2015, ApJ, 801, 44
\end{thebibliography}
\end{document}